\documentclass[14pt,epsf]{article}
\usepackage{graphicx}
\usepackage{epsf}
\begin{document}
\newcommand{\sheptitle}
{Estimations of baryon asymmetry  for  different neutrino mass models }
\newcommand{\shepauthor}
{Amal  Kr. Sarma $^{a,b}$,
  Hijam  Zeen  Devi$^{a}$  and  N. Nimai  Singh $^{a,}$
\footnote{Regular Associate, The Abdus Salam ICTP, Trieste, Italy\\
 E-mail: nimai03@yahoo.com}}
\newcommand{\shepaddress}
{$^{a}$Department of Physics, Gauhati University, Guwahati-781 014, India.\\
$^{b}$Department of Physics, D.R.College, Golaghat, Assam, India}
\newcommand{\shepabstract}
{We present a comparison  of   numerical predictions on  baryon 
asymmetry of the universe ($Y_{B}$) for different neutrino mass models.
 We start with  a very brief review on the main formalism of baryogenesis 
via leptogenesis through the  decay of heavy right-handed Majorana neutrinos,
and then  calculate the  baryon asymmetry of the universe for known
 six  neutrino mass models viz.,
 three quasi-degenerate, two inverted hierarchical and one 
normal hierarchical models, 
  which are  generated  from  canonical seesaw mechanism.  
The corresponding mass matrices for the right-handed Majorana 
neutrino $M_{RR}$  as well as the  
 Dirac neutrino $m_{LR}$  are fixed at the seesaw stage for 
generating correct light neutrino mass matrices $m_{LL}$ consistent with 
observations, and 
 are again employed in the calculation of baryogenesis. 
This two-tier procedure  removes possible 
 ambiguity on the choices of $m_{LR}$ and $M_{RR}$, and fixes the values of   
input parameters  at the seesaw stage. 
 We find that   the ranges of predictions from 
 both  normal hierarchical model (NHT3) 
and the degenerate model (DegT1A)
 having mass eigenvalues ($m_1, - m_2, m_3$), 
are almost consistent with  the observed baryon asymmetry of the universe.
Other two degenerate models (DegT1B, DegT1C) and one inverted hierarchical model 
(InvT2A) lead to very small baryon asymmetry, $Y_B \leq 10^{-19}$, whereas 
 inverted hierarchical model (InvT2B) gives 
larger $Y_B\geq 10^{-6}$.   
 Combining the present result with other predictions such as 
neutrino masses and mixings, and  stability  
 under radiative corrections in MSSM, 
the normal heirarchical  model appears to  be the most
 favorable choice of nature. Possible sources of uncertainty 
in the estimations are pointed out.}
\begin{titlepage}
\begin{flushright}
hep-ph/0604040
\end{flushright}
\begin{center}
{\large{\bf\sheptitle}} 
\bigskip\\
\shepauthor
\\
\mbox{}
\\
{\it\shepaddress}
\\
\vspace{.5in}
{\bf Abstract}\bigskip\end{center}\setcounter{page}{0}
\shepabstract
\end{titlepage}
\section{Introduction}
Left-right symmetric GUT models such as SO(10)GUT, assume the existence of 
heavy right-handed Majorana neutrinos, and also predict  small non-zero neutrino 
masses through the celebrated seesaw mechanism[1]. Various neutrino mass models 
describing neutrino mass patterns, have been proposed[2] within the framework of
 seesaw mechanism, and their predictions on 
neutrino masses and mixings have been studied thoroughly. In our earlier communication [2]
 we had studied the stability criteria 
of these models in the presence of left-handed triplet Higgs field in type-II seesaw formula, 
and also the stability  under radiative corrections in MSSM. 
The above  criteria had been  used to discriminate 
the neutrino mass models describing different patterns of neutrino masses. 
As a  continuation of our earlier effort in this direction, we are now interested 
in the present work to have a comparison of the different  theoretical 
predictions  of the  
cosmological  baryon asymmetry derived from  the above neutrino  mass models. This information 
may be treated as  an additional point in the process of  further  discrimination of
 neutrino mass models.
  
In SO(10)GUT, the   role of heavy right-handed Majorana neutrinos is manifold. In addition to its 
role in seesaw mechanism, it also plays an important role to explain  the observed  
baryon asymmetry of the  universe[3],
$$Y_{B}^{CMB}=(6.1^{+0.3}_{-0.2}) 10^{-10}.$$
Such baryon asymmetry can be dynamically generated if the particle interaction 
rate and the expansion rate of the universe satisfy  Sakharov's three famous conditions[4]:
(i) baryon non-conservation,  (ii) C and CP violation and (iii) departure from thermal equilibrium.
 Being  Majorana particles the heavy right-handed neutrinos automatically satisfy  
Sakharov's one of the above three conditions i.e., C and CP violation
 as they can have asymmetric decay to lepton and Higgs particles, and the processes 
occur at different rates for particles and antiparticles.
 Baryogenesis through leptogenesis by the electroweak sphaleron 
process has been widely   accepted as a correct approach   and a lot of work 
has been done in this direction[5,6,7,8,9,10].
The electroweak  sphaleron process violates baryon as well as  lepton number, and
 the  process is found to be  in equilibrium for a range of temperature i.e.,$100GeV\le T\le 10^{12}GeV$.

For a systematic estimation of baryon asymmetry, one has to correlate it  with the seesaw mechanism
 where both Dirac neutrino mass matrix and right-handed Majorana mass matrix are figured. 
In this context we refer to  our earlier work[2] where we had employed   the  
seesaw mechanism  to  successfully generate 
the  degenerate, inverted and normal hierarchical models of neutrino mass patterns,  
using the diagonal texture of $m_{LR}$ and non-diagonal texture of $M_{RR}$. 
In most of the  left-right symmetric SO(10)GUT models,  the Dirac neutrino mass matrix 
$m_{LR}$ can be either charged lepton mass matrix (referred to as case i) or
 up-quark mass matrix (referred to as case ii). 
However, we find that the  seesaw mechanism  alone fails to discriminate  the above 
two possible choices of  $m_{LR}$,  as well as the   correct choice of the neutrino 
mass model in question.
In the present  work, we will  estimate the baryon asymmetry of the universe
 using the same input values of the  parameters which were  already fixed  for the predictions of 
  masses and mixing angles at the seesaw stage. This in fact assures good
 predictions of neutrino masses and mixings.
 As emphasised before, such additional information on the calculations of the baryon asymmetry  
 may be used to further discriminate the two choices of 
$m_{LR}$, and also the  correct pattern of 
neutrino mass model in question.

 In section 2, we present a brief review  on the main formalism
 for estimating baryon asymmetry through the out-of-equilibrium decays of heavy 
right-handed neutrinos.   Section 3 presents the numerical calculations of the baryon
 asymmetry for  different neutrino mass models, 
and their comparison. Section 4 concludes with a summary and discussion.
Important expressions[2]  related to $m_{LL}$ and $M_{RR}$ for all  the neutrino mass models derived from 
seesaw formula, are relegated to  Appendix A for ready reference in the present calculations.
\section{A brief review on the  estimation of  baryon asymmetry}
\subsection{ Type-I seesaw formula and lepton asymmetry}
Type-I seesaw formula[1] relates the left-handed Majorana  neutrino mass matrix $m_{LL}$ 
and heavy right handed Majorana mass matrix $M_{RR}$ in a  simple way
\begin{equation}\label{eq:eps}
m_{LL} = -m_{LR}M_{RR}^{-1}m_{LR}^{T}
\end{equation}
where  $m_{LR}$ is the Dirac neutrino mass matrix. 
 In some left-right symmetric theories such as SO(10)GUT,
 the left-handed Higgs triplet $\Delta_L$ acquires a  vacuum expectation value
and this leads   two sources in the  neutrino mass matrix[11]:
$m_{LL} = m_{LL}^{I} + m_{LL}^{II},$
 where the first term is the usual seesaw term
 and the second one can be expressed as $m_{LL}^{II} = \gamma (M_W / v_R )^2 M_{RR}$. 
In general $\gamma$ is a function of various couplings, and without 
fine tuning  $\gamma$ is expected to be order of unity ( $\gamma\simeq 1$) . 
The modified seesaw formula (referred to  as Type-II) can be written as
\begin{equation}
m_{LL} = -m_{LR}M_{RR}^{-1}m_{LR}^{T} + \gamma (M_W / v_R )^2 M_{RR} 
\end{equation}
Three different possibilities  can  arise between the two  
contributing terms i.e., $m_{LL}^I$ and  $m_{LL}^{II}$. The situation where the second term is 
dominant over the first term, has the potential to
 connect [12,13,14,15] the large atmospheric neutrino mixing and b-$\tau$ unification 
in the context of minimal supersymmetric SO(10)models. In Type II seesaw formula, 
the lepton asymmetry $\epsilon_1$ can be written as:
\begin{equation}
\epsilon_1 = \epsilon^I + \epsilon^{II}
\end{equation}
 where  $\epsilon^I$ and  $\epsilon^{II}$ are the contributions of first term 
and second term of eq.(2) respectively.
For our calculation of lepton asymmetry we consider the model[5,6,7]
 where the asymmetric decay of the lightest of the heavy right-handed Majorana neutrinos,
 is assumed.  We confine our calculation  to the Type I contribution only, 
assuming the extra contribution is relatively very small($\gamma \sim 10^{-4}$).
 
The physical  Majorana neutrino $N_{R}$  decays into two modes:
$N_{R}\rightarrow l_{L}+\phi^{\dagger}$ \ \ and \ \ $\overline{l}_{L}+\phi,$
where  $l_{L}$ is the lepton and $\bar{l}_{L}$ is the antilepton. For CP-violating decay 
through  one-loop radiative correction by Higgs particle, the
 branching ratio for these two decay modes is  likely to be different. 
The CP-asymmetry which is caused by the intereference of tree level with one-loop corrections
 for the decays of lightest of heavy right-handed Majorana neutrino $N_{1}$ is defined by[6,8]
$\epsilon=\frac{\Gamma -\overline{\Gamma}}{\Gamma+\overline{\Gamma}}$
where  $\Gamma=\Gamma(N_{1}\rightarrow l_{L}\phi^{\dagger})$ 
and $\overline{\Gamma}=\Gamma(N_{1}\rightarrow \overline{l_{L}}\phi)$ are the decay rates.
A perturbative calculation from the interference between tree level and 
vertex plus self-energy diagrams for $N_1$ decay  gives[6,7,16]
\begin{equation}
\epsilon_{1} = \frac{1}{8\pi} \frac{\sum_{j}Im[(h h^{\dag})^2_{1j}]}{\sum_{i}|(h)_{1i}|^2} f(x_j)
\end{equation}
where Yukawa coupling of the Dirac neutrino mass matrix , is expressed by $h=m_{LR}/v$ , and 
$$f(x_{j})=f(\frac{M^{2}_{j}}{M_{1}^{2}})=f_{V}(x_{j})+f_{S}(x_{j})$$ 
with j=2,3. The function $f_{V}(x_{j})$ represents the vertex contribution
 and $f_{S}(x_{j})$ represents the 
self-energy contribution of the decay process of the right-handed Majorana neutrino.
 The value of $f(x_j)$ in MSSM is twice  its value for SM.
For  $x_{j}>>1$, the  expression (4) for SM case,  can be written in terms of neutrino mass matrix as[17,18,19,20]:
\begin{equation}
\epsilon^I = \epsilon_{1} = \frac{3 M_1}{16\pi v^2}\frac{Im[(h^\ast m^I_{LL}h^{\dag})_{11}]}
{(hh^{\dag})_{11}}
\end{equation}
For $x_{j}\simeq1$, we will have a very particular situation  where $M_{1}\simeq M_{2}<M_{3}$, 
 known as quasi-degenerate spectrum. The asymmetry is largely enhanced by a resonance
 factor which is given by[21,22,23]:
\begin{equation}
R= \frac{|M_1|}{2(|M_2|-|M_1|)}
\end{equation}
In such situation, we use the following expression[21]to calculate the lepton asymmetry,
\begin{equation}
\epsilon^I = -\frac{M_2}{8\pi v^2}\frac{Im[(h^\ast m^I_{LL}h^{\dag})_{11}]}{(hh^{\dag})_{11}} R
\end{equation}

\subsection{From lepton asymmetry to baryon asymmetry through sphaleron process}
The (B-L) conserving and (B+L) violating electroweak  sphaleron processes[5,6]
 give a set of  coupled [24] equations for the evolution of lepton number and baryon number.
 At temperature $T$ above the electroweak phase transition temperature (EWPT) $T_C$,
 the baryon asymmetry can be expressed in terms of (B - L) number density as [9]:
\begin{equation}
B(T>T_{c})=\frac{24+4N_{H}}{66+13N_{H}}(B-L)
\end{equation}
Again (B-L) asymmetry per unit entropy is just the nagative of $n_{L}/s$, since
 the baryon number is conserved in the right-handed Majorana neutrino decays.
At the electroweak phase transition temperature, any  primordial (B+L) will be washed out 
and the relation(8) can be written as[21,25]:
\begin{equation}
\frac{n_{B}}{s}\simeq -\frac{24+4N_{H}}{66+13N_{H}}\frac{n_{L}}{s}
\end{equation}
where $N_{H}$ is the number of Higgs doublet and  $n_{L}/s$  represents the lepton number
 to entropy ratio.
For   standard model(SM), $N_{H}=1$; and eq.(9) reduces to
\begin{equation}
\frac{n_{B}}{s}\simeq-\frac{28}{79}\frac{n_{L}}{s}
\end{equation}
For very slow B non-conserving interactions, the baryon to entropy ratio in a 
comoving volume is conserved. But the baryon to photon ratio does not remain 
constant due to the variation of photon density per comoving volume[8,26] 
at the different epoch of the expanding universe.
Considering the cosmic ray microwave background temperature $T\simeq2.7K$, 
we have $s=7.04 n_{\gamma}$. For SM  case we have 
$$(\frac{n_{L}}{s})^{SM}\simeq8.66\times10^{-3}\kappa_{1}\epsilon_{1}$$
where $\kappa_1$ is the dilution factor.
This leads to the observed baryon asymmetry  of the Universe [21],
\begin{equation}
Y_B^{SM}\equiv (\frac{n_{B}}{n_{\gamma}})^{SM} \simeq d\kappa_1\epsilon_1       
   = 0.0216 \kappa_1\epsilon_1
\end{equation}
 For our numerical estimation of baryon asymmetry for different neutrino mass patterns,
  we use the above  expression. 
\subsection{Out-of-Equilibrium Decay}
To get the resultant baryon asymmetry, an out-of-equilibrium situation has to be provided 
by the expanding universe. 
The desired condition is satisfied  if the temperature is smaller than the mass $M_{1}$ 
of the decaying neutrino at 
the time of its decay. The inverse decay is blocked[6,8,9] if the expansion rate 
is greater than the interaction 
rate $\Gamma$. Thus,
$$\Gamma<\frac{1.66\sqrt{g^{*}}T^{2}}{M_{pl}}$$
where $g^{\ast}$ is the effective numbers of degrees of freedom available at the
 temperature $T$ and $M_{pl}$ 
is the Planck mass scale. Substituting the expression of $\Gamma$ for the  decay of $M_{1}$, we have,
\begin{equation}
\frac{(hh^{\dag})_{11}M_{1}}{8\pi}<\frac{1.66\sqrt{g^{*}}T^{2}}{M_{pl}}
\end{equation}
For attaining out-of-equilibrium decay of $M_1$ we equate $T=M_{1}$, and we use a 
parameter[27] derived  from eq(12) as,
\begin{equation}
K\equiv(\frac{(hh^{\dag})_{11}M_{1}}{8\pi})(\frac{M_{pl}}{1.66\sqrt{g^{*}}M_{1}^{2}})<1
\end{equation}
How much the produced asymmetry is washed out is described by Boltzmann equation 
and can be parametrized by another parameter $\kappa$ known as dilution factor[8,27]:
\begin{equation}
\kappa_1\simeq\left\{\begin{array}{ll}
\frac{0.3}{K(lnK)^{0.6}} &\textrm{if $10\le K\le10^{6}$}\\
\frac{1}{2\sqrt{K^2+9}} & \textrm{if $0\le K\le10$}\\
\end{array}\right.
\end{equation}
An equivalent description for the  out-of-equilibrium decay can also be 
expressed by the decay parameter 
$$K=\tilde{m_{1}}/ m^\ast$$
where the effective neutrino mass is defined as 
$$\tilde{m_{1}}=\frac{(hh^{\dag})_{11}v^2}{M_{1}}$$
and the equilibrium neutrino mass is given by[28]:
\begin{equation}
m^\ast=\frac{16\pi^\frac{5}{2}}{3\sqrt{5}}g_\ast^\frac{1}{2}\frac{v^{2}}{M_{pl}}\simeq 1.08\times10^{-3}eV
\end{equation}
Here,  $g^\ast=106.75$, $M_{pl}=1.2\times10^{19}GeV$, $v = 174$GeV. 
Thus, for $\tilde{m_{1}}<m^\ast$, the heavy right-handed neutrino will satisfy the out-of-equilibrium
 condition. The values of K for different neutrino mass models 
under consideration, ranges from $3.7\times 10^2 $ to $10^{-1}$ 
as shown in table-3. Therefore,  the dilution factor $\kappa_1$ which takes into 
account the washout process due to inverse decays and lepton number violating 
scattering,  are calculated from expression(14). 
\section{Numerical calculation and results}
To start with, we predict  the left-handed Majorana  neutrino mass $m_{LL}$ using
 the seesaw formula[2]. As emphasised earlier, we adopt the diagonal form of Dirac neutrino
 mass matrix $m_{LR}$ for case(i) where the Dirac neutrino mass matrix is taken as 
charged lepton mass matrix and case(ii) where the Dirac neutrino mass matrix is taken as
 up-quark mass matrix. Using  the non-diagonal form of right-handed Majorana mass matrix 
$M_{RR}$ we obtain three patterns of  neutrino mass models: degenerate, inverted hierarchical and 
normal hierarchical. The corresponding expresions for $m_{LL}$ and $M_{RR}$ 
for these models, are given in Appendix-A for ready reference to the present calculation.

In the next step, the physical right-handed Majorana neutrino $N'$ is 
defined in the basis where $M_{RR}$ is diagonal with real positive eigen-values,i.e.,
\begin{equation}
N'=U_{R}^{T}N 
\end{equation}
where  $U_{R}^T$  is extracted from the relation,
\begin{equation}
M_{RR}^{diag}=U_{R}^{T}M_{RR}U_{R}=Diag(M_{1},M_{2},M_{3})
\end{equation}
This amounts to the re-definition of Dirac neutrino mass matrix $m_{LR}$:
\begin{equation}
m_{LR}\rightarrow m'_{LR}=m_{LR}U_{R}
\end{equation}
In general $U_R$ is complex, and the new $m'_{LR}$ in eq.(18) is made complex.
It can be emphasised that such transformation makes the seesaw term $m_{LL}$ invariant. 
In most calculations the Majorana CP phases enter through the construction of $m_{LL}$ 
but in the present case  such complex phases are originated from the diagonalising 
matrix $U_{R}$ of $M_{RR}$. However both approaches are equivalent as they are related 
by an inverse seesaw relation $M_{RR}= -m_{LR}m_{LL}^{-1}m_{LR}^T$, which is specially  true for 
diagonal form of mass matrix $m_{LR}$.
In our calculation  four  neutrino mass models given in appendix  viz.,  
 DegT1A, DegT1C, InvT2B, and NRT3,  
 have complex diagonalising matrix $U_R$ containing  complex Majorana phases. But for two models  
DegT1B and InvT2A, the  diagonalising  matrix $U_R$ is real, leading to  zero lepton asymmetry.
Since the non-zero lepton  asymmetry requires  complex Yukawa couplings
 i.e., complex  phases, the following  transformation is introduced by hand. Thus 
\begin{equation}
U_{R}\rightarrow U_{R}D
\end{equation}
where,
\begin{displaymath}
\mathbf{D}=\left(\begin{array}{ccc}
e^{i\alpha}&0&0\\
0&1&0\\
0&0&e^{i\beta}
\end{array}\right)
\end{displaymath}
contains  the complex Majorana phases. For further calculation of lepton asymmetry 
for the above two models we take  $\alpha = \beta = \pi/2$. 
It can be emphasised here  that such extra  phase transformation of $U_{R}$ does not change the seesaw 
term $m_{LL}$  and also  the term $(hh^{\dag})$ in eq.(4).

For demonstration  we first consider the normal hierarchical model as an example. 
The corresponding heavy right-handed neutrino mass matrix and light left-handed 
neutrino mass matrices  are collected from ref.[2] where we use the diagonal form 
of Dirac neutrino mass matrix $m_{LR}=Diag.(\lambda^m, \lambda^n, 1)v$. 
\underline{Normal Hierarchical Type3(NHT3)}:
\begin{displaymath}
\mathbf{M_{RR}}=\left(\begin{array}{ccc}
\lambda^{2m-1}&\lambda^{m+n-1}&\lambda^{m-1}\\
\lambda^{m+n-1}&\lambda^{m+n-2}&0\\
\lambda^{m-1}&0&1
\end{array}\right)v_{0}
\end{displaymath}
\begin{displaymath}
\mathbf{-m_{LL}^{I}}=\left(\begin{array}{ccc}
-\lambda^{4}&\lambda&\lambda^{3}\\
\lambda&1-\lambda&-1\\
\lambda^{3}&-1&1-\lambda^{3}
\end{array}\right)m_{0}
\end{displaymath}
We take the input values $\lambda=0.3$, $ m_0 = 0.03 eV$  leading  to 
 correct predictions of  neutrino mass parameters and mixing angles: 
\begin{center}
$\Delta m^2_{21}= 9.04\times 10^{-5}eV^2, \Delta m^2_{23}= 3.01\times10^{-3}eV^2$\\
$\tan^2\theta_{12}= 0.55,\sin^22\theta_{23}= 0.98, \sin\theta_{13}= 0.074$
\end{center}
These are almost consistent with recent data[2]. The predictions on 
neutrino masses and mixing angles  for all the models, are presented in Table-1.

\underline{For case(i)}: We use the charged lepton mass matrix,  $m_{E} =m_{LR}$,
 $(m,n)=(6,2)$,  $v_{0} = 1.01\times10^{15}$ GeV, $m_{0}$= 0.03 eV, 
 $m_{LR}=m_{E}=diag(\lambda^{6}, \lambda^{2},1)v$,with $v = 174$GeV.
This leads to 
\begin{center}
\begin{displaymath}
\mathbf{M_{RR}}=\left(\begin{array}{ccc}
1.7887\times10^9&2.2083\times10^{11}&2.4537\times10^{12}\\
2.2083\times10^{11}&7.3610\times10^{11}&0\\
2.4537\times10^{12}&0&1.0097\times10^{15}
\end{array}\right)
\end{displaymath}
\end{center}
For real and positive eigen values[29] of $M_{RR}$ the corresponding diagonalising matrix is  
\begin{center}
\begin{displaymath}
\mathbf{U_{R}}=\left(\begin{array}{ccc}
 0.964044 e^{-i\frac{\pi}{2}}    &  -0.265732    & 0.00242998\\
 -0.265732  e^{-i\frac{\pi}{2}}   &  -0.964047    & 5.31822\times10^{-7}\\
 -0.00234248  e^{-i\frac{\pi}{2}} & 0.000646238   & 0.999997
\end{array}\right)
\end{displaymath}
\end{center}
and the three heavy masses are 
\begin{center}
$M_{RR}^{diag}=diag( 6.51\times10^{10},7.97\times10^{11},1.01\times10^{15}).$
\end{center}
The Yukawa coupling `$h$' is defined[23]  by $h=m_{LR}/v$, where  $v$ is the
electroweak vacuum expectation value. 
 As the Dirac neutrino mass matrix is rotated
 in the basis where right-handed Majorana neutrino mass matrix is diagonal, the new
 Yukawa coupling becomes: $ h = m_{LR}U_{R}/v$
\begin{center}
\begin{displaymath}
\mathbf{h}=\left(\begin{array}{ccc}
-0.000702788  e^{-i \frac{\pi}{2}}  & -0.000193718    & 1.77146\times10^{-6}\\
0.0239159   e^{-i \frac{\pi}{2}}    & -0.0867642      & 4.7864\times10^{-8}\\
0.00234248  e^{-i \frac{\pi}{2}}    & 0.000646238     & 0.999997
\end{array}\right)
\end{displaymath}
\end{center}
\underline{For case(ii)}: We again use the up-quark mass matrix  $m_{up}=m_{LR}$,
  $(m,n)=(8,4)$, $v_{0} = 1.01\times10^{15}$GeV,  $m_{0} = 0.03$ eV ,
$m_{LR}=m_{up}=diag(\lambda^8,\lambda^4,1)v$. This leads to   
\begin{center}
\begin{displaymath}
\mathbf{M_{RR}}=\left(\begin{array}{ccc}
1.4495\times10^{7} & 1.7895\times10^{9} & 2.2092\times10^{11}\\
1.7895\times10^{9} & 5.9648\times10^{9} & 0\\
2.2092\times10^{11}& 0                  & 1.0102\times10^{15}
\end{array}\right)
\end{displaymath}
\end{center}
The corresponding diagonalising matrix is:
\begin{center}
\begin{displaymath}
\mathbf{U_{R}}=\left(\begin{array}{ccc}
0.964047e^{-i\frac{\pi}{2}}    & -0.265733  & 0.0002187\\
-0.265733e^{-i\frac{\pi}{2}}     & -0.964047  & -3.87423\times10^{-10}\\
-0.000210837  e^{-i\frac{\pi}{2}}& 0.0000581162 &-1
\end{array}\right)
\end{displaymath}
\end{center}
\begin{center}
$M_{RR}^{diag}=diag( 5.2699\times10^{8},6.4571\times10^{9},1.01\times10^{15})$
\end{center}
The Yukawa coupling `h' is:
\begin{center}
\begin{displaymath}
\mathbf{h}=\left(\begin{array}{ccc}
-7.02787\times10^{-4}e^{-i\frac{\pi}{2}}   & -1.93719\times10^{-4}&1.59431\times10^{-7}\\
0.02391 e^{-i\frac{\pi}{2}}   &-0.08676   & 3.4868\times10^{-11}\\
2.10837\times10^{-4} e^{-i\frac{\pi}{2}}  & 5.81161\times10^{-5}&1
\end{array}\right)
\end{displaymath}
\end{center}
To calculate the dilution factor first  we calculate the effective mass $\tilde{m}$ 
for each case  (Table-3). For normal hierarchical case we have for case(i):
$\tilde{m_{1}}=2.47\times10^{-4}$ eV
and for case(ii):
$\tilde{m_{1}}=2.47\times10^{-4}$ eV
The corresponding dilution factors are $0.17$ and $0.17$  for case(i) and case(ii) respectively.
Considering the decay of $N_1$, the baryon asymmetry is found to be:\\
for case(i) where $m_{LR}=m_{E}$,
 $$Y_{B}^{SM}=2.17\times10^{-9}$$
and for case(ii) where $m_{LR}=m_{up}$$,
 $$Y_{B}^{SM}=1.76\times10^{-11}$
 Following the same procedure we  calculate the 
baryon asymmetry for all the models given in Appendix-A. The results are presented in
 Tables 1-5. As emphasised earlier, we use only the values of the  parameters fixed at the first stage 
where seesaw formula operates, in the  estimation of  baryon asymmetry in the present calculation.
\small
\begin{table}[tbp]
\begin{center}
\begin{tabular}{cccccc}
\hline
Type&$\Delta m^{2}_{21}[10^{-5}eV^{2}]$&$\Delta m^{2}_{23}[10^{-3}eV^{2}]$&$\tan^{2}\theta_{12}$&$\sin^{2}2\theta_{23}$&$\sin\theta_{13}$\\
\hline
DegT1A&$8.80$&$2.83$&$0.98$&$1.0$&$0.0$\\
DegT1B&$7.91$&$2.50$&$0.27$&$1.0$&$0.0$\\
DegT1C&$7.91$&$2.50$&$0.27$&$1.0$&$0.0$\\
InvT2A&$8.36$&$2.50$&$0.44$&$1.0$&$0.0$\\
InvT2B&$9.30$&$2.50$&$0.98$&$1.0$&$0.0$\\
NHT3&$9.04$&$3.01$&$0.55$&$0.98$&$0.074$\\
\hline
\end{tabular}
\hfil
\caption{\footnotesize 
 Predicted values of the  solar and atmospheric neutrino mass-squared 
differences  and three mixing parameters 
calculated from $m_{LL}^{I}$ derived from  seesaw formula  in the Appendix A.}
\end{center}
\end{table}

\small
\begin{table}[tbp]
\begin{center}
\begin{tabular}{cc|c}
\hline
Type   &  Case(i):$|M_{j}|$            &  Case(ii):$|M_{j}|$\\           
\hline
DegT1A & $4.28\times 10^9,1.16\times10^{10},3.84\times10^{13}$  & $3.47\times10^7,9.42\times10^7,3.81\times10^{13}$\\
DegT1B & $4.05\times10^7,6.16\times10^{11},7.6\times10^{13}$  & $3.28\times10^5,4.98\times10^9,7.6\times10^{13}$\\
DegT1C & $4.05\times10^7,6.69\times10^{12},6.99\times10^{12}$  & $3.28\times10^5,4.85\times10^{11},7.81\times10^{11}$\\
InvT2A & $3.28\times10^{8},9.70\times10^{12},6.79\times 10^{16}$  & $2.64\times10^6,7.92\times10^{10},6.70\times10^{16}$\\
InvT2B & $5.6527\times10^{10},5.6532\times10^{10},5.38\times10^{16}$  & $4.5971\times10^8,4.5974\times10^8,5.34\times10^{16}$\\
NHT3  & $6.51\times10^{10},7.97\times10^{11},1.01\times10^{15}$  & $5.27\times10^8,6.45\times10^9,1.01\times10^{15}$\\
\hline
\end{tabular}
\hfil
\caption{\footnotesize
 The three right-handed Majorana neutrino masses in GeV  
for both case (i) and case (ii). The expressions of $M_{RR}$ are collected from Appendix A.}
\end{center}
\end{table}

\small
\begin{table}[tbp]
\begin{center}
\begin{tabular}{cccc|ccc}
\hline
 & For & case(i) & & For & case(ii)&\\
\hline
Type   &$  \tilde{m}_1$ &$ K$ &$ \kappa_1$ & $  \tilde{m}_1$ &$ K$ &$ \kappa_1$ \\             \hline
DegT1A & $3.76\times10^{-3}$&$3.48 $&$ 0.11$ & $3.74\times10^{-3}$&$3.46$&$0.11$\\
DegT1B & $0.40$&$370$& $2.79\times 10^{-4}$              & $0.40$&$370$&$ 2.79\times 10^{-4}$\\
DegT1C & $0.40$&$370$&$2.79\times 10^{-4}$               & $0.40$&$370$&$ 2.79\times 10^{-4}$\\
InvT2A & $0.05$&$46.3$&$ 2.9\times 10^{-3}$              & $0.05$&$46.3$&$2.9\times 10^{-3}$\\
InvT2B & $2.85\times10^{-4}$&$0.26$&$ 0.17 $& $2.83\times10^{-4}$&$0.26$&$0.17$\\
NHT3   & $2.47\times10^{-4}$&$0.23$&$0.17$  & $2.47\times10^{-4}$&$0.23$&$0.17$\\
\hline
\end{tabular}
\hfil
\caption{\footnotesize 
Calculation of values of effective mass parameter $\tilde{m_{1}}$ in eV, and dilution factor $\kappa_1$.}
\end{center}
\end{table}

\small
\begin{table}[tbp]
\begin{center}
\begin{tabular}{ccc|cc}
\hline
Type  &  $\epsilon^I$(case i)     & $\epsilon^I$(case ii)  & $Y_{B}^{SM}Case(i)$ &  $Y_{B}^{SM}Case(ii)$ \\ 
\hline
DegT1A & $2.10\times10^{-6}$    & $1.71\times10^{-8}$  & $4.99\times10^{-9}$ & $4.06\times10^{-11}$ \\
DegT1B & $2.66\times10^{-18}$   & $2.16\times10^{-20}$ & $1.60\times10^{-23}$ & $1.30\times10^{-25}$ \\
DegT1C & $1.74\times10^{-18}$   & $1.69\times10^{-20}$ & $1.05\times10^{-23}$ & $1.02\times10^{-25}$ \\
InvT2A & $1.59\times10^{-14}$   & $1.27\times10^{-16}$ & $9.94\times10^{-19}$ & $7.96\times10^{-21}$ \\
InvT2B & $1.47\times10^{-2}$    & $1.62\times10^{-4}$  & $5.40\times10^{-5}$  & $5.94\times10^{-7}$  \\
NRT3   & $5.90\times10^{-7}$    & $4.78\times10^{-9}$  & $2.17\times10^{-9}$  & $1.76\times10^{-11}$ \\
\hline
\end{tabular}
\hfil
\caption{\footnotesize
 Calculation of lepton asymmetry $\epsilon^I$  and baryon asymmetry $Y_{B}$ for case(i) and 
 case(ii) for respective neutrino mass models given in  Appendix-A.}
\end{center}
\end{table}
\small
\begin{table}[tbp]
\begin{center}
\begin{tabular}{c|cccc}
\hline
$(m,n)$  & $(8,4)$ &  $(6,2)$  &  $(4,2)$   &$(7,2)$\\
\hline
$M_1(GeV)$&$5.27\times10^{8}$&$6.51\times10^{10}$&$1.16\times10^{12}$&$1.87\times10^{10}$\\ 
$\epsilon_1$ & $4.78\times 10^{-9}$ & $5.9\times10^{-7}$ & $1\times10^{-5}$ & $1.7\times10^{-7}$\\
$ Y_B^{SM}$   & $1.76\times10^{-11}$ & $2.17\times10^{-9}$ & $3.24\times10^{-8}$ & $6.24\times10^{-10}$\\
\hline
\end{tabular}
\hfil
\caption{\footnotesize
Calculation of baryon asymmetry for normal hierarchical mass model  for different choices of $(m,n)$ pair.}
\end{center}
\end{table} 


\section{Summary and Discussion}
We summarise the main points in this work. We have attempted to give a 
comparison of the numerical estimations on baryon asymmetry of the universe 
 for different neutrino
 mass models.  We have investigated   these  neutrino mass models 
(degenerate, inverted hierarchical and  normal hierarchical) at two consecutive stages:
(i) from neutrino oscillation parameters (Table 1) using seesaw mechanism  and 
(ii) from the observed baryon asymmetry (Table 4),  within
 a unified framework. In a model where baryon asymmetry is explained via lepton asymmetry 
through sphaleron process, the out-of-equilibrium decay of the right-handed Majorana neutrino plays a 
 decesive role.  As the out-of-equilibrium decay of the lightest heavy Majorana neutrino $M_{1}$ 
is essential for fruitful
 leptogenesis, we consider the effective mass parameter $\tilde{m_{1}}$ for
 all the models presented in Table 3 where  only inverted hierarchical model(InvT2B)  
and normal hierarchical model (NRT3) 
satisfy the out-of-equilibrium decay condition. In these two models the 
effective neutrino mass is less
 than the equilibrium neutrino mass $m^{\ast}(\simeq10^{-3})$ i.e., $\tilde{m_{1}}<m^\ast$.

We present the numerical predictions on  solar and atmospheric  neutrino mass-squared differences 
and three mixing angles, which are calculated using seesaw formula in Table 1.
All the input parameters in $M_{RR}$ and $m_{LR}$ are fixed at this stage. Table 1 indicates 
good predictions  for normal hierarchical model and inverted hierarchical model( InvT2A).
  However, the inverted hierarchical 
(Type 2A) model  is not stable under radiative corrections in MSSM unlike normal
 hierarchical model(NRT3). In Table 2 we present the masses of the 
physical right-handed 
Majorana neutrinos, which are extracted through the diagonalisation of $M_{RR}$. 
 Further, the calculated values of baryon asymmetry for all neutrino mass models
 are shown in Table 4 which clearly shows  the competitive 
nature of degenerate (Type IA) model [30] and Normal hierarchical model. 
For both  models, the produced baryon asymmetry  is slightly
 lesser than the experimental value in  case (ii) and slightly larger in  case (i).
This possibly implies that the actual form of Dirac neutrino mass matrix may lie 
between these two extreme cases. A  result of our investigations in this regards  for normal hierarchical 
mass model, is presented in Table-5 for different values of $( m, n)$ pair in $m_{LR}$ [31]. 
As the  mass  of the lightest right-handed Majorana neutrino  $M_{1}$ is concerned,
 the  model   agrees with the famous  Ibarra-Davidson  bound [32]
 i.e.,$M_1 > 4\times 10^8$ GeV.
  Considering the 
stability criteria  under radiative corrections in MSSM as well as  good predictions on
 oscillation mass paramerters and mixing 
angles, the normal hierarchical  model is quite favourable.
 Future neutrino oscillation experiments may   confirm 
this conjecture. Finally we also point out  some ambiguities on the choice of the value of 
coefficent $d= 0.216$. The present value is found to be  $2.2$
 times larger than the value appeared in refs.[16,19,23,33] and  $0.74$ times 
smaller than the value appeared  in ref.[34].

\section*{Appendix A}
Here we collect  the various right-handed Majorana mass matrices $M_{RR}$
 and the light left-handed neutrino mass matrix $m_{LL}$ from ref.[2] for 
ready reference in the text. Here we use diagonal form of 
Dirac neutrino mass matrix $m_{LR}= Diag (\lambda^m, \lambda^n, 1)v$. 
The nomenclature of the different neutrino mass 
models has been changed here.
\begin{flushleft}
\underline{Degenerate Type1A(DegT1A) with mass eigenvalues $(m_1, -m_2, m_3)$:}
\end{flushleft}
\begin{center}
\begin{displaymath}
\mathbf{M_{RR}}=\left(\begin{array}{ccc}
 -2\delta_{2}\lambda^{2m}&(\frac{1}{\sqrt{2}}+\delta_{1})\lambda^{m+n}&(\frac{1}{\sqrt{2}}+\delta_{1})\lambda^{m}\\
(\frac{1}{\sqrt{2}}+\delta_{1})\lambda^{m+n}&(1/2+\delta_{1}-\delta_{2})\lambda^{2n}&(-1/2+\delta_{1}-\delta_{2})\lambda^{n}\\
(\frac{1}{\sqrt{2}}+\delta_{1})\lambda^{m}&(-1/2+\delta_{1}-\delta_{2})\lambda^{n}&(-1/2+\delta_{1}-\delta_{2})
\end{array}\right)v_{0}
\end{displaymath}
\end{center}
\begin{center}
\begin{displaymath}
\mathbf{-m_{LL}^{I}}=\left(\begin{array}{ccc}
 (-2\delta_{1}+2\delta_{2})&(\frac{1}{\sqrt{2}}-\delta_{1})&(\frac{1}{\sqrt{2}}-\delta_{1})\\
(\frac{1}{\sqrt{2}}-\delta_{1})&(1/2+\delta_{2})&(-1/2+\delta_{2})\\
(\frac{1}{\sqrt{2}}-\delta_{1})&(-1/2+\delta_{2})&(1/2+\delta_{2})
\end{array}\right)m_{0}
\end{displaymath}

\end{center}
Here, $\delta_{1}=0.0061875 ,\delta_{2}=0.0031625 ,\lambda=0.3, m_{0}=0.4$eV and $v_0 = 7.57\times 10^{13}$GeV\\
\begin{flushleft}
\underline{Degenerate Type1B(DegT1B) with mass eigenvalues $(m_1, m_2, m_3)$}:
\end{flushleft}
\begin{center}
\begin{displaymath}
\mathbf{M_{RR}}=\left(\begin{array}{ccc}
(1+2\delta_{1}+2\delta_{2})\lambda^{2m}&\delta_{1}\lambda^{m+n}&\delta_{1}\lambda^{m}\\
\delta_{1}\lambda^{m+n}&(1+\delta_{2}\lambda^{2n}&\delta_{2}\lambda^{n}\\\delta_{1}\lambda^{m}&\delta_{2}\lambda^{n}&(1+\delta_{2})
\end{array}\right)v_{0}
\end{displaymath}
\end{center}
\begin{center}
\begin{displaymath}
\mathbf{-m_{LL}^{I}}=\left(\begin{array}{ccc}
(1-2\delta_{1}-2\delta_{2})&-\delta_{1}&-\delta_{1}\\
-\delta_{1}&(1-\delta_{2})&-\delta_{2}\\
-\delta_{1}&-\delta_{2}&(1-\delta_{2})
\end{array}\right)m_{0}
\end{displaymath}
\end{center}
Here, $\delta_{1}=7.2\times10^{-5},\delta_{2}=3.9\times10^{-3},\lambda=0.3,m_{0}=0.4$eV and $v_0 = 7.57\times10^{13}$GeV.

\begin{flushleft}
\underline{Degenerate Type1C(DegT1C) with mass eigenvalues $(m_1, m_2, -m_3)$}:
\end{flushleft}
\begin{center}
\begin{displaymath}
\mathbf{M_{RR}}=\left(\begin{array}{ccc}
(1+2\delta_{1}+2\delta_{2})\lambda^{2m}&\delta_{1}\lambda^{m+n}&\delta_{1}\lambda^{m}\\

\delta_{1}\lambda^{m+n}&\delta_{2}\lambda^{2n}&(1+\delta_{2})\lambda^{n}\\\delta_{1}\lambda^{m}&(1+\delta_{2})\lambda^{n}&\delta_{2}
\end{array}\right)v_{0}
\end{displaymath}
\end{center}
\begin{center}
\begin{displaymath}
\mathbf{-m_{LL}^{I}}=\left(\begin{array}{ccc}
(1-2\delta_{1}-2\delta_{2})&-\delta_{1}&-\delta_{1}\\
-\delta_{1}&-\delta_{2}&(1-\delta_{2})\\
-\delta_{1}&(1-\delta_{2})&-\delta_{2}
\end{array}\right)m_{0}
\end{displaymath}
\end{center}
Here, $\delta_{1}=7.2\times10^{-5},\delta_{2}=3.9\times10^{-3},\lambda=0.3,m_{0}=0.4$eV and $v_0 = 7.57\times10^{13}$GeV.
\begin{flushleft}
\underline{Inverted Hierarchical Type2A(InvT2A) with mass eigenvalues $(m_1, m_2, m_3)$}:
\end{flushleft}
\begin{center}
\begin{displaymath}
\mathbf{M_{RR}}=\left(\begin{array}{ccc}
\eta(1+2\epsilon)\lambda^{2m}&\eta\epsilon\lambda^{m+n}&\eta\epsilon\lambda^{m}\\
\eta\epsilon\lambda^{m+n}&1/2\lambda^{2n}&-(1/2-\eta)\lambda^{n}\\
\eta\epsilon\lambda^{m}&-(1/2-\eta)\lambda^{n}&1/2
\end{array}\right)\frac{v_{0}}{\eta}
\end{displaymath}
\end{center}
\begin{center}
\begin{displaymath}
\mathbf{-m_{LL}^{I}}=\left(\begin{array}{ccc}
(1-2\epsilon)&-\epsilon&-\epsilon\\
-\epsilon&\frac{1}{2}&(\frac{1}{2}-\eta)\\
-\epsilon&(\frac{1}{2}-\eta)&\frac{1}{2}
\end{array}\right)m_{0}
\end{displaymath}
\end{center}
Here, $\eta=0.0045 ,\epsilon=0.0055,\lambda=0.3 ,m_{0}=0.05$eV and $v_0 = 6.06\times 10^{14}$GeV
\begin{flushleft}
\underline{Inverted Hierarchical Type2B(InvT2B) with mass eigenvalues $(m_1, -m_2, m_3)$}:
\end{flushleft}
\begin{center}
\begin{displaymath}
\mathbf{M_{RR}}=\left(\begin{array}
{ccc}
\lambda^{2m+7}&\lambda^{m+n+4}&\lambda^{m+4}\\
\lambda^{m+n+4}&\lambda^{2n}&-\lambda^{n}\\
\lambda^{m+4}&-\lambda^{n}&1
\end{array}\right)v_{0}
\end{displaymath}
\end{center}
\begin{center}
\begin{displaymath}
\mathbf{-m_{LL}^{I}}=\left(\begin{array}{ccc}
0&1&1\\
1&-(\lambda^{3}-\lambda^{4})/2&-(\lambda^{3}+\lambda^{4})/2\\
1&-(\lambda^{3}+\lambda^{4})/2&-(\lambda^{3}-\lambda^{4})/2
\end{array}\right)m_{0}
\end{displaymath}
\end{center}
Here,$\lambda=0.3 , m_{0}=0.035$eV and $v_0 = 5.34\times 10^{16}$GeV

\begin{flushleft}
\underline{Normal Hierarchical Type3(NHT3) with mass eigenvalues $(m_1, m_2, m_3)$}:
\end{flushleft}
\begin{center}
\begin{displaymath}
\mathbf{M_{RR}}=\left(\begin{array}{ccc}
\lambda^{2m-1}&\lambda^{m+n-1}&\lambda^{m-1}\\
\lambda^{m+n-1}&\lambda^{m+n-2}&0\\
\lambda^{m-1}&0&1
\end{array}\right)v_{0}
\end{displaymath}
\end{center}
\begin{center}
\begin{displaymath}
\mathbf{-m_{LL}^{I}}=\left(\begin{array}{ccc}
-\lambda^{4}&\lambda&\lambda^{3}\\
\lambda&1-\lambda&-1\\
\lambda^{3}&-1&1-\lambda^{3}
\end{array}\right)m_{0}
\end{displaymath}
\end{center}
Here,$\lambda=0.3,m_{0}=0.03$eV and $v_0 = 1.01\times 10^{15}$GeV.
\section*{Acknowledgements}
One of us (AKS) would like to thank  Prof. Utpal Sarkar for useful communication.
 NNS wishes to acknowledge
fruitful discussions with Prof.K.S.Babu and  Prof. E.J.Chun during WHEPP-9 at Institute of Physics,
Bhubaneswar, India.

\end{document}